# Short-Packet Interleaver against Impulse Interference in Practical Industrial Environments

Ming Zhan, Member, IEEE, Zhibo Pang, Senior Member, IEEE, Dacfey Dzung, Life Member, IEEE,
Kan Yu, Member, IEEE, and Ming Xiao, Senior Member, IEEE

M. Zhan is with the College of Electronics and Information Engineering, Taizhou University, Taizhou 318000, Zhejiang, China (e-mail: zmdjs@swu.edu.cn).
Z. Pang is with the ABB Corporate Research, Forskargr¨and 7, SE-721 78 Vasterås, Sweden, he is also a visiting professor with the KTH Royal Institute of Technology, 10044 Stockholm, Sweden (e-mail: pang.zhibo@se.abb.com, zhibo@kth.se).
D. Dzung is with the ABB Corporate Research, 5405 Baden, Switzerland (e-mail: dacfey.dzung@ch.abb.com).
K. Yu is with the Department of Computer Science and Information Technology, School of Engineering and Mathematical Sciences, La Trobe University, Bendigo Campus, Edwards Rd, Flora Hill VIC 3552, Australia (e-mail: kan.yu@hotmail.com).
M. Xiao is with the School of Electrical Engineering, KTH Royal Institute of Technology, 10044 Stockholm, Sweden (e-mail: mingx@kth.se).

*Abstract*—The most common cause of transmission failure in Wireless High Performance (WirelessHP) target industry environments is impulse interference. As interleavers are commonly used to improve the reliability on the Orthogonal Frequency Division Multiplexing (OFDM) symbol level for long packet transmission, this paper considers the feasibility of applying short-packet bit interleaving to enhance the impulse/burst interference resisting capability on both OFDM symbol and frame level. Using the Universal Software Radio Peripherals (USRP) and PC hardware platform, the Packet Error Rate (PER) performance of interleaved coded short-packet transmission with Convolutional Codes (CC), Reed-Solomon codes (RS) and RS+CC concatenated codes are tested and analyzed. Applying the IEEE 1613 standard for impulse interference generation, extensive PER tests of CC(1/2) and RS(31, 21)+CC(1/2) concatenated codes are performed. With practical experiments, we prove the effectiveness of bit interleaved coded short-packet transmission in real factory environments. We also investigate how PER performance depends on the interleavers, codes and impulse interference power and frequency.

## I. Introduction

Over the past decade, with the rapid advances in emerging technologies (e.g. information and communication technology, sensors, intelligent robots), the traditional manufacturing industry has been undergoing a digital transformation at a tremendous pace [1]. As some industrial wireless communication standards, for example WirelessHART, WIA-PA and WIA-FA [2], [3], [4], can match the moderate ms level latency requirement in typical Process Automation (PA) and Factory Automation (FA) applications, they are not applicable in some critical fields of industrial applications, such as Power Systems Automation (PSA) and Power Electronics Control (PEC), ultra-high reliability, extremely low latency and determinism are strictly required [5], [6], [7]. For transmission of 100 bits short-packet with lower than 10 us latency, the Wireless High-Performance (WirelessHP) was proposed as a solution to meet the stringent requirements [8]. However, as a common cause of transmission failure in factory environments, impulse interference is an obstacle in achieving the goal [9]. Therefore, the IEEE 1613 standard stipulates that wireless communication systems for control purposes should be designed to resist impulse interference [10], [11], [12].

In industrial application scenarios where WirelessHP is applied, the positions of central controller, sensors and actuators are usually fixed [8], [13]. However, the wireless channel in industrial environments is significantly different from civil home or office environments, as human/robot movement, equipment shape and location, other wireless communication systems working at the same frequency band, and electromagnetic radiation from arcs and welding, may create a complex wireless environment. When channel coding can achieve $10^{-7}$ Packet Error Rate (PER) in factory environments [14], we still found the PER performance is drastically degraded when suffering from impulse interference when Orthogonal Frequency Division Multiplexing (OFDM) is employed. Unfortunately, impulse interference cannot be eliminated as it comes from the industrial production processes, for example switching on/off of high voltage equipment, or from arcs in electric welding. Another source is the communication system itself, e.g. spurious emissions from the power supply and the ground wire (refer to Section IV-B). To improve impulse interference resisting capability, relocating positions of error bits is a solution [15], [16], [17]. There have been many literature that emphasize bit or symbol interleaved coded transmission [18], [19], [20], [21]. However, there is an apparent lack of studies conducted in real factory environments, only preliminary results are presented in [22], [23]. More importantly, one of the key differences between the WirelessHP system and current industrial wireless communication standards is that the physical layer of the former is specially designed for short-packet transmission (refer to Section II-C). As a result of the new physical layer structure, we are motivated to investigate the possibility of adopting interleaver to the WirelessHP paradigm.

For this purpose, the WirelessHP physical layer protocols were implemented on a Universal Software Radio Peripherals (USRP) hardware platform, and experiments with different parameter combinations of coding schemes, code rates and interleaver types were performed in factory environments. The PER assessment uses four types of interleavers, namely, two

The work was supported by the National Nature Science Foundation of China under Grant 61671390. (Corresponding author: Zhibo Pang.)

block interleavers (one for packet and another for symbol), the 3rd Generation Partnership Project (3GPP) turbo code interleaver, and the S-random interleaver. Different from conventional technique that the interleaving/deinterleaving process is performed on one OFDM symbol for long packet, this research extends interleavers to a wider OFDM frame level for short packet. Referring to the IEEE 1613 standard [11], we generate impulse interference to simulate electromagnetic interference from other devices and analyze resilience of interleaved coded short-packet transmission. We show interleavers can improve the PER performance to a certain degree for different coding schemes. To the best of our knowledge, this is the first research on interleaver in the physical layer with single preamble for packet detection and 100 bits short packet. The contributions of this research are summarized below.

1) By practical experiments in real factory environments, we prove that bit interleaved coded transmission is an effective solution to improve the PER performance of WirelessHP transmission subject to impulse interference. Among all the factors in our experiments, the impulse interference power dominates in affecting PER performance, followed by the impulse frequency and interleaver structure. To fully exploit the advantages of interleaver for short-packet transmission, this paper presents the experimental setup and guidance.

2) For short-packet coded transmission, we show that there is room for optimization of the interleaver structure and matching coding schemes to enhance impulse interference resisting capability. Under low-energy impulse noise conditions, interleaving on the OFDM symbol level can slightly improve PER performance but is inferior to interleaving on the frame level. This shows better PER performance can be achieved by interleaving the packet bits in a wider frame range. We also show that coding schemes and interleaver structure have different tolerances to power and frequency of impulse interference. This suggests further research in this field.

The rest of this paper is organized as follows. Section II gives background knowledge of WirelessHP and related works on interleaved coded transmission. In Section III, we present the construction of interleavers from a theoretical perspective. The USRP-PC experimental configuration and the generation of impulse interference are detailed in Section IV. PER comparison and performance analysis of interleavers in coded short-packet transmission in factory environments are detailed in Section V. At last, conclusions are summarized in Section VI.

## II. BACKGROUND KNOWLEDGE AND RELATED WORKS

### A. Impulse noise models

The impulse noise phenomenon, which is sequences of impulses with different duration, intensities and individually occurs in random time, is first proposed by Middleton in [24]. Then in [25], the statistical noise models for the man-made and natural impulse noise were developed, among which the most widely used is the Middleton Class A impulse noise model. Sources distribution of the impulse noise is a Poisson process with parameter A (also termed as the impulse index), and the Probability Density Function (PDF) of a generated noise sample follows the Gaussian distribution with variance $\sigma_i^2 / A$ ($\sigma_i^2$ is the impulse noise power). In power line communication systems, the impulse noise effects can also be represented by the Bernoulli-Gauss model [26]. A noise sample $m_k$ is expressed as $m_k = w_k + i_k$, where $w_k$ is the background Additive White Gaussian Noise (AWGN) with mean zero and variance $\sigma_w^2$, and $i_k = b_k \times g_k$ is the impulse noise, where $b_k$ is a Bernoulli process that take values in $\{0,1\}$ with probability $P(b_k = 1) = p$, $g_k$ is white Gaussian random sample with zero mean and $\sigma_i^2$. In case parameter $A$ of Middleton Class A noise model is sufficiently low, the sum of its first few terms are accurate enough, and the Bernoulli-Gauss model is considered an accurate approximation of the Middleton Class A noise model. In [27], the generation of impulse noise with appropriate characteristics of amplitude, length, inter-arrival and spectrum are comprehensively detailed.

Wireless channels in industrial scenarios significantly differ from home/office environments. Impulse noise may be generated from frequently switching operations to the power source, which is typically non-Gaussian [28]. The shape, amplitude, duration, and arrival time of an impulse is unpredictable, making it is complicated to model the PDF [29]. Moreover, the models of impulse noise may vary significantly for different locations. An impulse is identified if the amplitude or power of the received signal is higher than predefined threshold [30]. [31] and [32] proposed to model the impulse noise shape using sinc function. By categorizing the impulse noise into five general classes [33], periodic impulsive noise (50 KHz to 200 KHz) is introduced with the shape similar to the damped sinusoid [34]. To evaluate the residual error rate of a safety critical communication system in the worst industrial environments, [35] stipulates 5 KHz and 100 KHz impulse for 15 ms and 0.75 ms duration of testing, respectively. In the IEEE 1613 standard [10], [11], [12], the Surge Withstand Capability (SWC) test is required for wireless communication devices employed in electrical substations.

### B. OFDM impulse noise cancellation algorithms

OFDM is a high-speed data communication technology that data are modulated on a number of subcarriers and transmitted in parallel. The deleterious effect of fading channel is spread over the whole OFDM symbol rather than completely corrupted few adjacent bits. Considering the impulse noise has very short duration that the spectral components are spread over all subcarriers, the long OFDM symbol is less sensitive to the impulse noise in a bursty channel. However, in case that the impulse noise energy exceeds a certain threshold, the OFDM system performance may be seriously degraded [26], [36]. In [37], an algorithm is proposed to estimate the impulse shape for adaptive noise cancellation. This scheme is further developed in [30], by using the redundancy in the guard band frequency domain, the impulse noise corrupted samples are recalculated. Based on the pilot and null subcarriers, the similarity between the Reed-Solomon (RS) code in the complex field and Fast Fourier Transform (FFT) is applied to remove the impulse noise samples in the OFDM symbol [38], [39]. In [40], the authors proposed to represent the RS code with filterbank, and merged it with the OFDM modulation to form an improved RS-OFDM scheme for effective impulse noise mitigation. The work in [41] proposed to use channel estimation pilot subcarriers to cancel the impulse noise, then this approach is enhanced for time-frequency selective channels [42]. In [36] and [43], the blanking and clipping nonlinearities were proposed to compress impulse noise. In the blanking nonlinearity algorithm, an impulse is detected if the amplitude exceeds a threshold, and the impulse noise is deleted by replacing the sample value with zeros. While in the clipping nonlinearity algorithm, the impulse noise is removed by amplitude truncation. Lately, an optimal threshold selection algorithm is addressed in [44]. Based on preliminary estimated noisy signal, [45] and [46] proposed to estimate the impulse noise component in each of the received sample, and subtract the noise from the samples if the estimated value is large enough. In [47], a joint time domain/frequency domain reduction scheme is used to combat the adverse effect of impulse noise. At the cost of increased complexity, [31] and [48] proposed to apply the impulse noise suppression algorithm iteratively. In [49], the iterative impulse noise suppression strategy is extended to a fast convergence variant. Other effective mitigation algorithms include the null-subcarriers based impulse noise position detection technique in [50], the iteratively performed adaptive blanking threshold scheme in [51], the adaptive infinite impulse response notch filter based periodic impulse noise suppression algorithm in [34], and the sparse signal reconstruction algorithm based impulse noise estimation and cancellation scheme in [52].

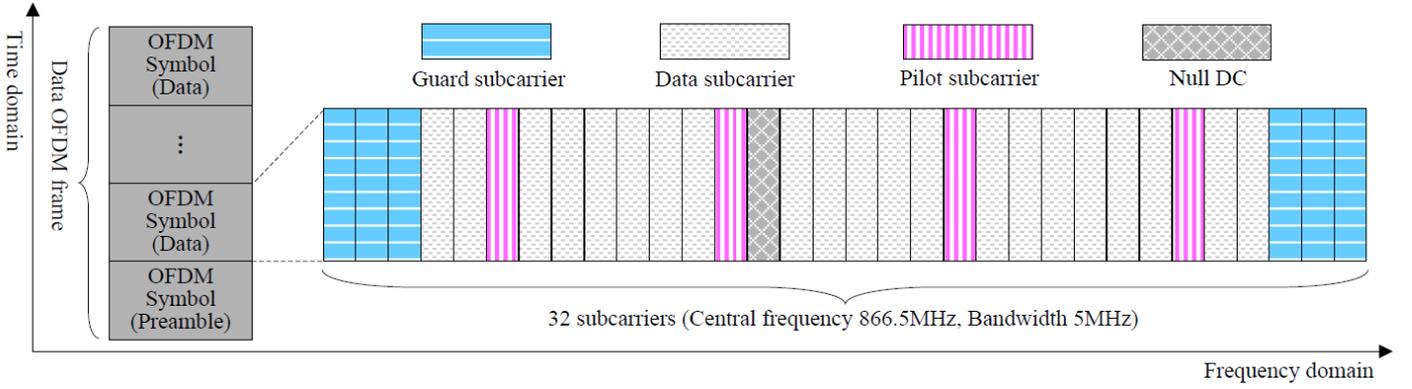

Fig. 1. Structure of WirelessHP data OFDM frame in time/frequency domain.

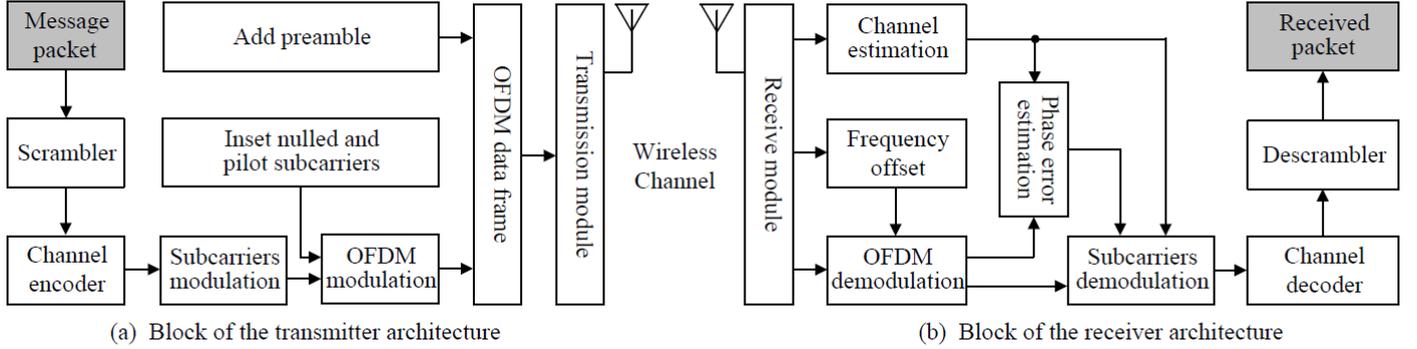

Fig. 2. Block diagrams of WirelessHP transmitter and receiver.

*C. WirelessHP and its developments*

In the existing industrial wireless communication standards, WirelessHART, ISA100.11a and WIA-PA are the modified versions of the IEEE 802.15.4 standard, while WIA-FA is originated from the IEEE 802.11g/n standard [7]. As these standards facilitate their easier deployment and faster popularization, they are not suitable for the low latency requirement in PSA and PEC for critical applications. Because the physical layer of the IEEE 802.15.4 and IEEE 802.11g/n are designed with long preambles for long packet transmission, neither of these technologies is optimal for achieving ultra-low latency in short-packet transmission. A new paradigm, WirelessHP, is proposed to alleviate the efficiency problems with short-packet transceiving by adopting innovative solutions at its physical layer [8]: i) In order to enable fast packet detection, synchronization, and channel estimation, one OFDM symbol is used as a preamble. With static positions of transmitter and receiver nodes and short communication distances, such a short preamble suffices. It is also assumed that fixed Time Division Multiple Access (TDMA) is used with the predefined cyclic transmission, so that the controller and the sensor/actuator nodes know at which time slot a data OFDM frame is to be transmitted or received. ii) Packet length is minimized. As network (star) topology and location of the central controller, sensors and actuators are static, most network parameters, such as the number of nodes and long addresses, need not be transmitted in each packet. Other parameters for physical layer settings, like the channel coding schemes, code rate and packet length, are set in the network initialization stage and thus can also be omitted from the packet overhead, thus minimizing the packet length.

Figure 1 illustrates the WirelessHP data OFDM frame structure in time and frequency domains. For a coded packet length of some 102 bits, one data OFDM frame is composed of one OFDM preamble and a number of data OFDM symbols. By comprehensive parameters optimization to meet the extremely low latency requirement [8], a 5 MHz frequency band is evenly divided into 32 subcarriers, among which 4, 1, and 6 subcarriers are devoted to pilot, nulled direct current and guard subcarriers, while the remaining 21 subcarriers are used for data transmission. The number of data OFDM symbols is determined by four factors: the message packet length, the modulation order, the code rates, and the number of data subcarriers. Figure 2 explains the block diagrams of WirelessHP transmitter and receiver. At the transmitter side, the message packet is first processed by the scrambler, then the coded packet is evenly divided into $N_{data}^{sym}$ parts (seen Eq. (7)), and bits in each part are assigned to the data subcarriers for modulation. By using FFT, the pilot, nulled direct current, guard and data subcarriers are generated to form data OFDM symbols. Subsequently, the preamble and data OFDM symbols are assembled to form one data OFDM frame. In the transmission module, the OFDM data frame is modulated to a central frequency for transmission. At the receiver side, the preamble OFDM symbol in the received frame is extracted for synchronization, frequency offset and channel estimation, which are then used to demodulate the data OFDM symbols. In the data OFDM symbols, the pilot subcarriers are extracted to correct the phase error for more accurate data subcarriers demodulation. After being processed by the channel decoder and descrambler, the received packet is used for data refresh. It is worthy pointing out that the OFDM impulse noise cancellation algorithms reviewed in Section II-B are effective solutions for long packet transmission with long preambles. In case of the WirelessHP system, the data OFDM frame is designed for short-packet transmission with only one OFDM symbol as the preamble, it would be a long-term research to develop an effective impulse noise cancellation algorithm for complicated industrial wireless channels through hardware experiments. In the current WirelessHP version, the impulse noise cancellation techniques are not employed.

Based on practical experiments, [8] demonstrates the effectiveness of the WirelessHP physical layer protocols. Furthermore, it shows the packet transmission duration of 100 bits message packet can be shortened to 1.6 us if a 160 MHz bandwidth is used. [53] proposed a fast packet detection algorithm with a single OFDM symbol as the preamble. By adopting differential detection and prediction mechanisms, a packet detection error rate in the order of $10^{-6}$ is verified. Channel coding is proven to be an effective way to improve PER when incorporated into the WirelessHP physical layer, with only slight increase in transmission latency [54]. [14] explores the influence of modulation order and transmission efficiency on coded short-packet transmission, this work is the first to report that $10^{-7}$ PER is achievable in real factory environments.

*D. Interleaved coded transmission*

The full potential of interleavers to improve transmission reliability was realized with the invention of turbo codes in [55]. To improve the PER performance in a fading channel, [56] presents a novel mechanism by interleaving the coded bits before been modulated for wireless transmission. Based on the analysis in [57], channel coding can improve performance if the channel interleaver is long enough to take advantage of the diversity effect. However, for applications requiring a high degree of delay-insensitivity, such as industrial wireless control, where control message packets are usually short, the entire transmitted packet may experience the same interference if the transmitted packet duration is shorter than the interference duration [58]. The resulting block fading effect thus cannot provide sufficient diversity.

Shannon limit approaching coding schemes, for example the turbo and Low Density Parity Check (LDPC) code families, have been incorporated with interleaver for purpose of improved error correction capability [59], [60], as this simple and traditionally used technique is an effective solution for long packet. But these codes are not superior to traditional convolutional/block codes when used for short-packet transmission [61]. With today's parallel decoding architectures, multi-Gbps throughput decoders are easily realized for turbo and LDPC codes [62], [63]. However, high throughput channel code decoder is the precondition for industrial wireless control, while ultra-high reliability and low latency are the most emphasized metrics. Considering the high error floor of turbo and LDPC codes for short-packet coded transmission, and the iterative decoding algorithms originated long latency, turbo and LDPC codes are less preferred for industrial wireless control. The study of bit interleaving and traditional codes, for example Convolutional Codes (CC), Reed-Solomon codes (RS), is conducted to improve performance for scenarios in which latency and reliability are both essential [58]. Most work concerns theoretical analyses using idealized channel models. Unfortunately, these assumptions cannot reflect the fundamental characteristics in industrial environments. The interleaved coded transmission tests in [22] were performed in a factory, but the physical layer structures of the employed wireless communication standards (IEEE 802.15.4 and IEEE 802.11) are quite different from the WirelessHP system considered in this work.

III. INTERLEAVERS FOR THE WIRELESSHP PHYSICAL LAYER

As reviewed in Section II-B and C, in OFDM the data is mapped on a number of subcarriers, and then these data are transmitted in parallel by using FFT. With the longer OFDM symbol duration, the impulse noise spectrum is spread over all of the subcarriers, thus OFDM is robust to the impulse noise with low energy. As the impulse noise energy is only spread in the range of one OFDM symbol, the superiority may lose in case of high-energy impulse noise. A possible solution is to incorporate interleaver into the WirelessHP physical layer. Resorting to an interleaver at the transmitter side, bits in the coded short-packet are relocated within the whole data OFDM frame. Viewed form the time domain at the receiver side, the high-power impulse noise impaired samples in one OFDM symbol are spread in more OFDM symbols after deinterleaving. In this paper, we explore the feasibility of applying bit

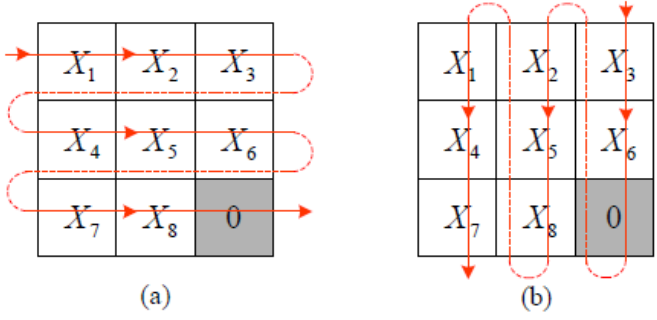

Fig. 3. An example of the packet block interleaver: (a) writing packet bits $X_1X_2X_3X_4X_5X_6X_7X_8$ into the 3×3 matrix, (b) reading interleaved packet bits $X_3X_6 0 X_2X_5X_8X_1X_4X_7$ out of the 3×3 matrix (0 is the padded bit).

interleaving in the WirelessHP physical layer with the following unique characteristics.

1) Only one OFDM symbol is employed on the WirelessHP physical layer as the preamble, then fast synchronization and channel estimation algorithms are employed for packet detection and OFDM demodulation. While these techniques can significantly reduce transmission latency, the effect on demodulation and decoding performance should be considered, because multi-path fading and interference from other wireless devices working in the same band are important factors which may impair synchronization and channel estimation.

2) Industrial control applications require $10^{-6}$ to $10^{-9}$ level of PER to guarantee stable operation [64]. Resorting to lower code rate, [14] has shown $10^{-7}$ PER is achievable, and also found long impulse duration will cause severe PER degradation. As OFDM can withstand the weak surge impulse with short duration, the PER performance deteriorates rapidly when the impulse noise frequency increases. Although high energy or high frequency impulse noise may represent the worst cases in industrial scenarios and the probability is low, the PER degradation should be seriously addressed if ultra-high reliability is definitely required.

3) In industrial applications targeted by WirelessHP, impulse interference represents the main type of interference leading to burst errors in the received packet. Especially, the impulse duration is much shorter than that of the data OFDM symbols. To achieve ultra-high reliability while keeping the latency overhead at an acceptable level for 100 bits short-packet transmission, redistributing the position of the corrupted bits within the packet is a preferred option. This has not yet been addressed for WirelessHP and related literature.

To investigate the effectiveness of interleavers in the WirelessHP physical layer, four kinds of interleavers are evaluated. Note that $L_2$ and $N_{data}^{sym}$ are explained in Section IV-A.

**1) Packet block interleaver** - By assuming $M$ and $N$ as the number of rows and columns respectively, a packet block interleaver can be modeled by a $M \times N$ matrix. First, all $L_2$ bits in a coded packet are written into the $M \times N$ matrix row by row. Then, the interleaved packet is output by reading out bits from the matrix column by column. It is worth pointing out that the bits can be written into and read out in different manners [65]. In our experiments, bits in the coded packet are written into each row of the matrix from the left side to the right side, then start from the right column, the interleaved bits are read out from the top to the bottom from the right column, as illustrated in Figure 3 with an example of $L_2 = 8$ bits. $M$ and $N$ are the matrix parameters calculated by Eq. (1) and (2), respectively. The value of $L_3$, i.e. the number of zero bits padded to fit the interleaver length, is given by Eq. (3).

$$M = \lfloor \sqrt{L_2} \rfloor + 1 \qquad (1)$$

$$N = \begin{cases} M-1, & \text{if } M(M-1) \geq L_2 \\ M, & \text{if } M(M-1) < L_2 \end{cases} \qquad (2)$$

$$L_3 = M \times N - L_2 \qquad (3)$$

**2) Symbol block interleaver** - Different from the packet block interleaver, where bits are redistributed in the range of the whole data OFDM frame, in a symbol block interleaver, the padded bits and coded bits are partitioned into $N_{data}^{sym}$ parts, each corresponding to one OFDM symbol, and interleaving is performed within each part. Since each part can be interleaved/deinterleaved in parallel, processing latency can be significantly reduced as compared to packet block interleaving. In this research, Binary Phase Shift Keying (BPSK) is employed for modulation, and each OFDM symbol includes 21 data subcarriers, each part contains 21 bits. In the same way shown in Figure 3, the 21 bits are interleaved by using a $3 \times 7$ matrix.

**3) 3GPP interleaver** - As a Quadratic Polynomial Permutation (QPP) interleaver designed for LTE-Advanced turbo codes [66], the 3GPP interleaver reduces the correlation between adjacent bits with low hardware overhead. Assuming $Y = y_0, y_1, \cdots, y_I, \cdots, y_{K-1}$ as a coded packet, $Y' = y'_0, y'_1, \cdots, y'_I, \cdots, y'_{K-1}$ as the interleaved packet, where $K$ is the packet length, the interleaved bit $Y'_i$ is corresponding to the bit $Y_{\pi(i)}$ in the coded packet $Y$, and $\pi(i)$ is computed by Eq. (4).

$$\pi(i) = (f_1 \cdot i + f_1 \cdot i^2) \bmod K \qquad (4)$$

where $f_1$ and $f_2$ are the interleaving coefficients for a packet length $K$ that predefined by the LTE-Advanced standard [67]. For example if $K = 120$, $f_1$ is 103 and $f_2$ is 90. To fit the coded packet length $L_2$, $K$ is chosen as follows: i) $K$ is greater or equals to $L_2$. ii) The number of padded zeros bits $L_3 = K - L_2$ should be as small as possible.

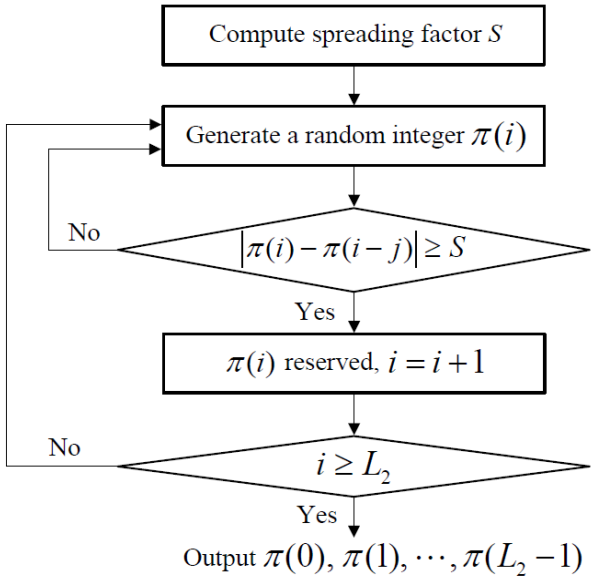

Fig. 4. Flow diagram of S-random interleaver generation.

4) *S-random interleaver* - Assuming $S$ as the spreading factor of a S-random interleaver, for a $L_2$-length coded packet, the spreading factor $S$ is defined as $S = \lfloor \sqrt{L_2 / 2} \rfloor$. Given a spreading factor $S$, we can compute the interleaver $\pi(i)$ sequence as follows ($i = 0, 1, \cdots, L_2 - 1$): i) Get an integer $\pi(i)$ in $[1, L_2]$ by using a random function generator. ii) Calculate the absolute difference between $\pi(i)$ and $\pi(i-j)$, where $j = 1, 2, \cdots, S$. If all values of the absolute difference are more than $S$, $\pi(i)$ is reserved and step i) is repeated to generate the next integer $\pi(i+1)$. Otherwise, the above two steps are repeated until a qualified value $\pi(i)$ is obtained. iii) When all integer values of the interleaver sequence $\pi(0), \pi(1), \cdots, \pi(i), \cdots, \pi(L_2 - 1)$ are successfully generated, the S-random interleaver sequence is output for packet interleaving. A flow diagram of S-random interleaver generation is shown in Figure 4, refer to [68] for more details of S-random interleaver.

For example if CC(2/3) is used for $L = 100$ bits coded transmission, by Eq. (5) and (6), $L_1$ and $L_2$ is calculated as 2 and 153, respectively. In case that the packet block interleaver is applied, we can calculated that $M = 13$, $N = 12$ and $L_3 = 3$ using Eq. (1), (2) and (3). Considering that $L_2 + L_3 = 156$, the number of data OFDM symbols is $N_{data}^{sym} = 8$ based on Eq. (7). Consequently, $L_4 = 12$ zero bits are padded to the interleaved 156-bit packet for OFDM modulation by Eq. (8).

IV. EXPERIMENTAL SETUP

*A. Hardware deployment of the experiment*

According to Figure 5 (a), an interleaving module is placed between the channel encoder and the OFDM modulation module, enabling the coded bits to be interleaved for wireless transmission. In comparison, Figure 5 (b) illustrates the receiver diagram, where the deinterleaving module is involved within a single OFDM symbol (for the symbol block interleaver) or within multiple OFDM symbols (for the other three interleavers) to disperse bit error bursts caused by time-domain impulse interference and/or frequency-domain narrowband interference. This should increase the possibility of recovering those erroneous bits by the error correction algorithm. Hardware deployment of the experiment is also presented in Figure 5, two Ettus USRPs X310 are deployed with a Line-of-Sight (LOS) distance of 10 meters. Both USRPs function as the transmitter and receiver, and both are connected to a PC with a Gigabit Ethernet cable to form the USRP-PC hardware testing platform. The practical scenario is illustrated in Figure 6: The experiments were conducted in an operating automated tea production workshop with a space of 42.1, 9.6 and 3.75 meters for length, width and height, respectively. Both USRP systems are perched on trolleys, and are configured to set up a LOS test scenario. Antennas used are W1910 penta band stubby models for an 866.5 MHz central frequency and a bandwidth of 5 MHz. Considering a typical industrial environment is shared by many wireless devices, the transmission power of each device is limited to avoid mutual interference. In this research, the effective transmission power is approximately 18 dBm, since the transmission power for USRP X310 is 17 dBm when the transmission gain is set to 15 [69], and the peak antenna gain is 1 dB at the frequency of 866.5 MHz.

As reviewed in Section II-C, WirelessHP mainly concentrates on improving the short-packet transmission e_ciency. On the physical layer, the numbers of guard, pilot and data subcarriers have been optimized, while the subcarrier modulation order parameter is optional [8]. In a noisy factory environment, [14] presented a comprehensive investigation of the PER performance for di_erent modulation orders. With the same data OFDM frame duration, distance and transmission power, BPSK is superior to Quadrature Phase Shift Keying (QPSK) and 8 Phase Shift Keying (8PSK) in term of PER. Although high order modulation can avail the advantage of interleaving burst error bits for long packet transmission, it may not suitable for industrial wireless control applications, where the message packet is short and the transmission power should be limited within an acceptable level. Therefore, BPSK is used in this research. As shown in Figure 5 (a), the 100 bits message packet is processed as follows before arriving at the OFDM modulation module:

1) At the transmitter USRP-PC side, $L_1$ zero bits are padded to the $L$ data bits to adjust the packet length for

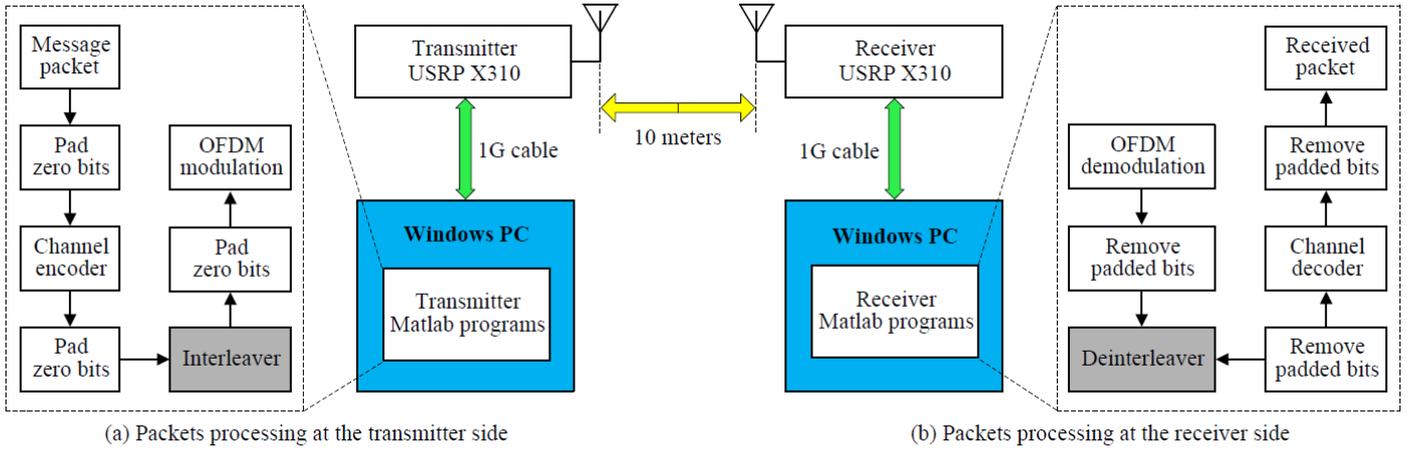

(a) Packets processing at the transmitter side  (b) Packets processing at the receiver side

Fig. 5. Diagrams of bit interleaved coded transmission in WirelessHP physical layer.

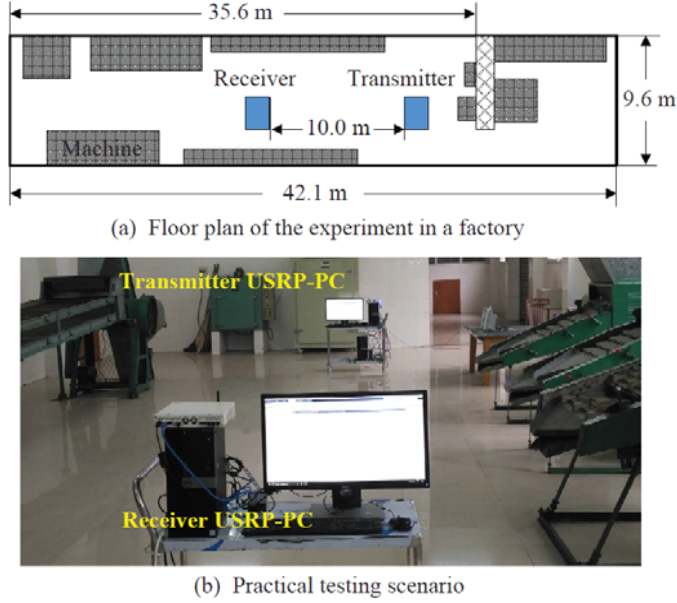

Fig. 6. Testing scenario in a real factory environment.

encoding, where $L_1$ and the coded packet length $L_2$ is computed by Eq. (5) and (6), respectively. Note that the code rate R is an irreducible fraction, and $R_o$ is the numerator of $R$.

$$L_1 = \left\lceil \frac{L}{R_o} \right\rceil \times R_o - L \quad (5)$$

$$L_2 = \frac{L + L_1}{R} \quad (6)$$

2) Each coded packet is padded with $L_3$ zero bits, if necessary, to fit for the interleaver length. Given that BPSK is used for the subcarriers modulation and each data OFDM symbol has 21 subcarriers for data transmission, the number of data OFDM symbols $N_{data}^{sym}$ can be computed by Eq. (7).

Since $N_{data}^{sym}$ is an integer, $L_4$ zero bits should be padded to the interleaved packet for OFDM modulation, which is given by Eq. (8).

$$N_{data}^{sym} = \left\lceil \frac{L_2 + L_3}{21} \right\rceil \quad (7)$$

$$L_4 = N_{data}^{sym} \times 21 - (L_2 + L_3) \quad (8)$$

As shown in Figure 5 (b), a reverse procedure is performed at the receiver USRP-PC side. After OFDM demodulation, deinterleaving and decoding, the corresponding zero bits are removed for further processing. For purpose of real-time industrial wireless control, the correctness of a decoded packet should be seriously checked. Considering the high packet refresh rate in critical applications, a corrupted packet may lose its significance even if this packet is retransmitted, as this packet arrives too late that cannot indicate the latest status of the perceived parameter. From the specific application point of view, reliability evaluation of industrial wireless control is based on PER, while bit error rate and symbol error rate are less significant. Consequently, a packet is considered in error, if one or more bits in the packet are incorrect, then the ratio of error packets to overall packets is computed to get PER performance.

*B. Generation of impulse interference waveforms*

In the IEEE 1613 standard [11], a series of clauses address the impulse interference testing for communications devices, such as Ethernet hubs, switches and routers installed in power transmission and distribution facilities. These include the SWC test parameters of impulse waveforms, the testing chamber and the environment setup. In specified scenarios, the tested devices are exposed to impulse voltage shocks in the order of thousands of volts. According to observed and measured data in actual substation installations, the impulse waveforms are modeled by parameters in Table I.

Switching power supply systems are another vital sources of impulse interference. Featured with small size, low cost, and high conversion efficiency, switching techniques are widely used in converters for industrial applications [70]. To increase power density and efficiency, the switching frequency should be increased. Limited by the rising switching loss at higher switching frequency, common Si-Insulated Gate Bipolar Transistor (IGBT) high power inverters are operated at under 20 KHz frequency [71], while state-of-the-art Si power Metal-Oxide- Semiconductor Field-Effect Transistor (MOSFET) converters operate at less than 200 KHz in low power applications [72]. With recent progress in semiconductor material and manufacturing technologies, the higher breakdown voltage and lower on-resistance advantages are exploited to develop high-switching frequency power components. In [73], the gallium nitride (GaN) switches are employed to realize a 65 W flyback converter working at 340 KHz with peak efficiency of about 90%. An improved prototype, i.e., the quasi-resonant converter, is proposed in [74] with 12V/3A output and switching frequency of 1 MHz. By using silicon carbide (SiC) MOSFET as the core switching component, [71] has built a 20 kW grid inverter working at 300 KHz, and Se-Hong Park et al. went further by increasing the switching frequency to 4 MHz for a 10 kW inverter [75]. When the wireless communication system gets power from the same source with the electrical equipment, the high frequency switching power introduced impulse interference can break into the baseband through the ground wire, the power supply system originated impulse interference may interfere with the transmitted data frame waveforms [76], [77]. Moreover, in case that an independent power supply is used for wireless communication system, non-linear effects may then cause the surge and switching pulses to generate impulse interference in the radio frequency bands where the wireless communication system operates. Therefore, investigating the impact of high-frequency switching power supply systems on WirelessHP is an important topic in industrial applications. The 2.5 KHz frequency value in Table II was specified by IEEE 1613 standard in 2013 and does not reflect the increased switching frequency in today's converters. Hence this paper considers impulse interference frequency up to 700 KHz.

TABLE I   Impulse interference waveform parameters of IEEE 1613 standard SWC tests

| Waveform polarity | Magnitude | Impulse duration | Frequency | Testing duration |
|---|---|---|---|---|
| Positive & Negative | 4 Kilovolts Tolerance ±10% | 50 ns Tolerance ±10% 50% value | 2.5 KHz Tolerance ±20% | ≥ 1 minute |

The impulse interference is modeled by referring to the relevant clauses in the IEEE 1613 standard. For impulse interference with frequency ranges from 50 to 700 KHz, the impulse is rectangular and bipolar, while the duration is chosen as 100 ns, since this assumption is considered typical in a factory environment [11], [12]. Regarding the impulse power, $\Gamma$ is defined as the ratio of the impulse peak power ($P_{ip}$) to the average power of the WirelessHP data OFDM frame ($P_{tp}$), and is calculated by Eq. (9) as below.

$$\Gamma = 10 \times \log_{10}^{P_{ip}/P_{tp}} \quad (dB) \qquad (9)$$

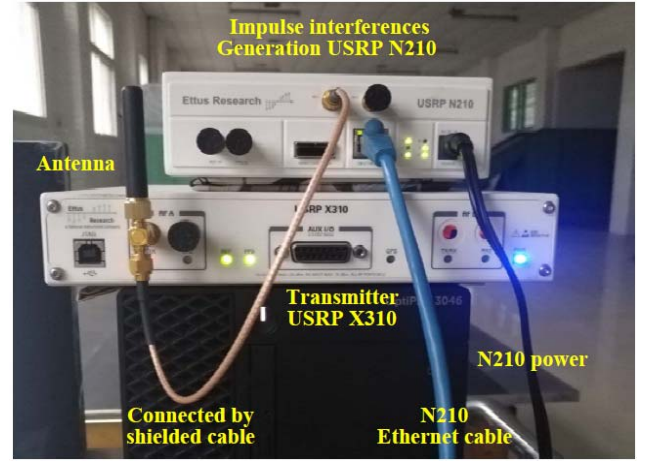

Fig. 7. Connection of impulse interference generator USRP N210 to transmitter USRP X310.

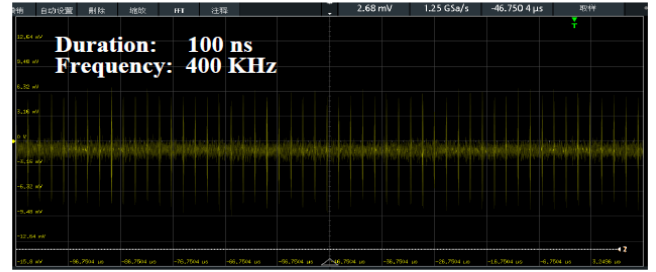

(a) Waveform of impulse interference

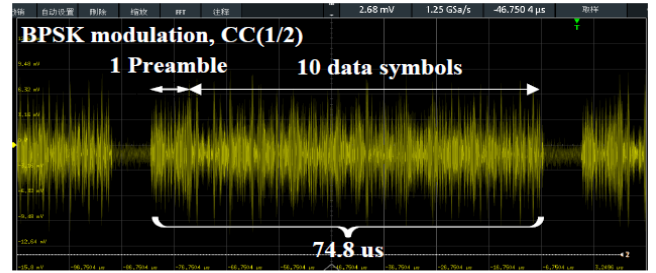

(b) Waveform of data OFDM frames

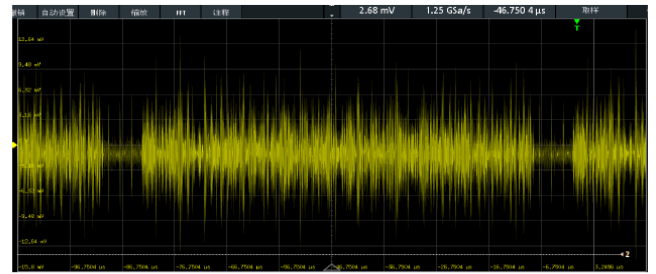

(c) Waveform of data OFDM frames and added impulse interference

Fig. 8. Example of waveforms.

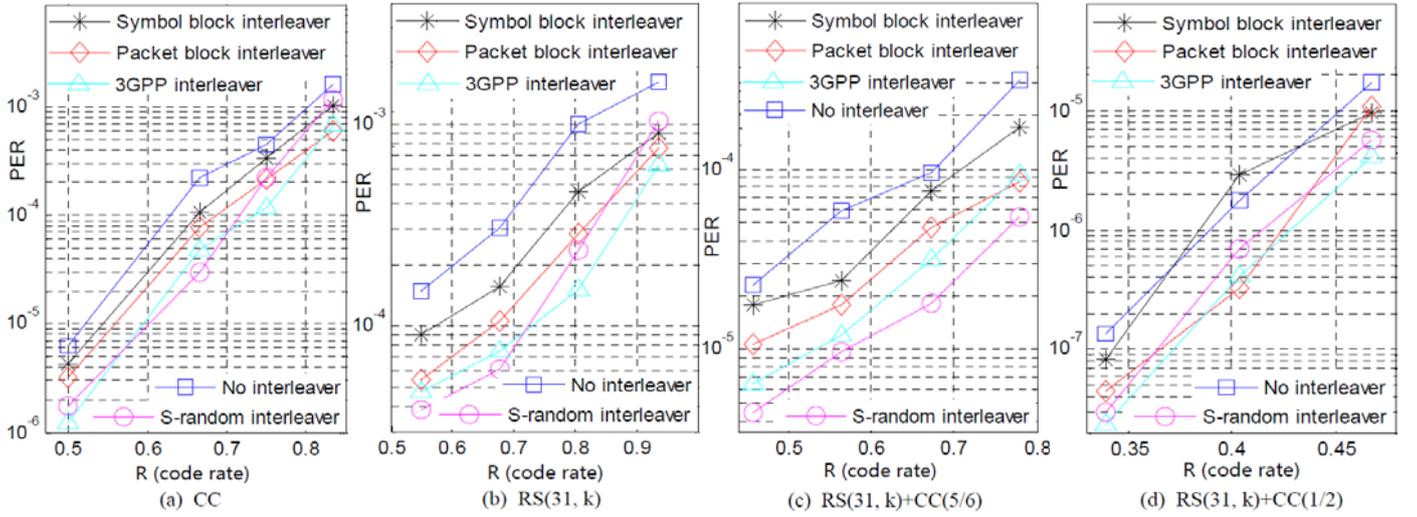

Fig. 9. PER performance comparison against different interleavers ($k = 29, 25, 21, 17$ for (b) and (c), and $k = 29, 25, 21$ for (d)).

In the impulse interference resisting capability testing, the transmitting USRP X310 has the same setup as described in Section IV-A. A Matlab program was designed to generate impulse interference waveforms by a USRP N210, which then modulates these waveforms to the central radio frequency of 866.5 MHz. This impulse interference signal is added to the transmitter antenna via a three-way connector as shown in Figure 7. By adjusting the transmitting gain of the USRP N210, the value of $\Gamma$ is set for four levels for practical experiments. An example of the generated impulse interference waveform, the transmitted packet waveform, and their combined waveform is shown in Figure 8. It is worthy pointing out that, based on the USRP parameters setup, duration of each OFDM symbol is 6.8 us, the data OFDM frame in Figure 8 (b) has 11 symbols with the whole duration of 74.8 us.

## V. EVALUATION OF INTERLEAVERS IN WIRELESSHP PHYSICAL LAYER FOR SHORT-PACKET TRANSMISSION

Four types of interleaver are implemented on the USRP-PC platform. CC, RS and RS+CC concatenated codes are applied to generate bit interleaved packets for OFDM transmission. In the receiver Matlab program, the decoding algorithm for CC and RS codes is the hard decision Viterbi and Berlekamp-Massey algorithm, respectively. A set of experiments, lasting over seven months, were conducted in real factory environments. This section first analyzes the effect of interleavers on PER performance for coded packet. Then the resilience of interleavers with these codes is investigated. Note that in practice, PER depends on the actual factory environments where the wireless communication system is deployed. In this research, we show how PER is affected by different interleavers and codes, for given impulse interference frequency and power.

### A. PER comparison among different interleavers

Three channel coding schemes are adopted in our PER performance evaluation. Since convolutional codes are superior to block codes in terms of latency, the first scheme is CC with the constraint length 7 and generator polynomial [171 133] in octal. The second coding scheme is the RS(31, k) code, as RS codes have powerful burst error correction capability. For a 100-bit level short-packet, the symbol length m of the RS code is 5 bits. Because longer symbol length would require more padded zero bits for the RS encoder, and shorter symbol length would mean weaker burst error correction capability. As the third coding scheme, we use the RS(31, k)+CC concatenated codes, where RS and CC are the outer and inner codes, respectively.

Figure 9 indicates the PER performance improvements by applying four types of interleavers in the testing scenario shown in Figure 6. As presented in these figures, in most cases lower PERs are achieved by using these interleavers. Compared with cases when packet block interleaver, symbol block interleaver and no interleaver are applied, PER performance of the 3GPP and S-random interleavers is much better. This may be because the 3GPP and S-random interleavers disperse adjacent bits to a wider distance. For example, when the S-random interleaver is employed, the position of neighbor bits are moved to a minimal distance of S bits. Thus interleaving is an effective solution to improve the reliability for short-packet transmission in factory environments.

Among the four interleavers, the symbol block interleaver achieves the slightest PER performance improvement. This can be explained by the fact that the symbol block interleaver moves bit position only within 21-bit symbols, while in the other interleavers bits are interleaved within the whole data OFDM frame. However, the PER with symbol block interleaver is still lower than that without any interleavers in most cases. According to [22], the PER performance may be degraded by using interleaver in Additive White Gaussian Noise (AWGN) channel, so our results may indicate that the channel in the testing environments has narrowband interference. Furthermore, considering that the packet block interleaver, 3GPP interleaver and S-random interleaver achieve better PER improvement than the symbol block interleaver, it indicates that impulse interference crossing multiple OFDM symbols causes burst errors in the received packet. Hence, interleaving performed in the time domain helps to resist impulse interference.

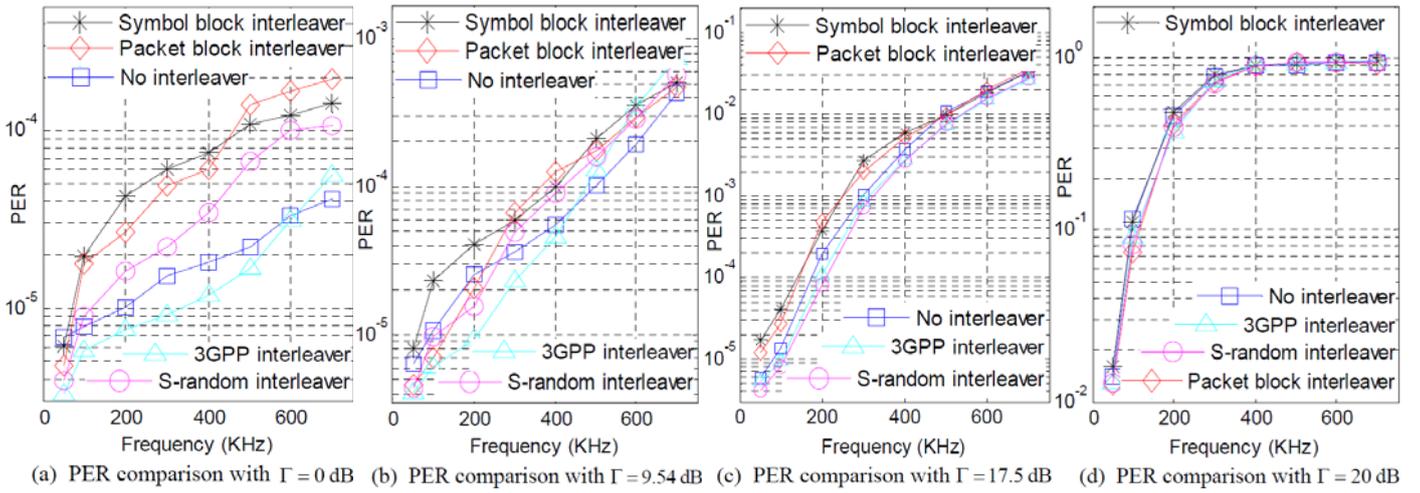

Fig. 10. PER comparison of impulse interference with different power ratio $\Gamma$ for CC(1/2).

*B. Extreme PER performance testing against high frequency impulse interference*

The experimental results of Figure 9 do not give a systematic analysis of the effect of interleavers on PER performance. Hence in this section the performance of the interleavers against impulse interference is investigated using impulse interference generated by a USRP N210. To explore the extreme performance, the interference power ratio $\Gamma$ is selected as 0, 9.54, 17.5 and 20 dB, as these values cover a range where the PER changes from slightly degraded to complete failure. CC(1/2) and RS(31, 21)+CC(1/2) concatenated codes are selected for this investigation, since they outperform the other codes in the high reliability region. For CC(1/2) coded transmission, the results are presented in Figure 10. Power ratio $\Gamma$ and impulse frequency are the key factors that affect PER performance. The power of the impulse interference depends on the emission power of the interference and the propagation mechanism (in particular the distance) from the interference source to the wireless receiver antenna. It should be pointed out that higher impulse frequency increases the number of impulses per data OFDM frame, which means higher interference energy (at given interference peak power) affecting the transmitted packets. As seen from Figure 10 (a), (b) and (c), when the impulse frequency is 50 KHz, the results show that using interleaver can improve the PER performance, and the PERs are only slightly degraded by the power ratio $\Gamma$. This phenomenon means when only a small number of impulse waveforms are included in a data OFDM frame, the interleaved coded strategy still can correct the corrupted bits on condition of lower/moderated impulse interference power. Moreover, when the PERs of Figure 10 (a), (b) and (c) at 50 KHz are compared with that of Figure 9 (a) at the code rate of 1/2, we find the PER values are very similar for the tested interleavers and no interleaver case. As the impulse frequency increase, the testing results show that using interleaver is not preferred to improve the reliability. The reason can be explained as follows: due to the excessive number of impulses in a short-packet, the impulse interference power and duration are higher. Consequently, the deinterleaving process cannot disperse all the burst bits, and may likely generate new burst errors in a short range. In Figure 10 (d) where $\Gamma$ is increased to 20 dB, the PERs are drastically descended to the range of $1.3 \times 10^{-2} \sim 1$. This phenomenon reflects the fact that the WirelessHP transmitting/receiving mechanism, to a certain degree, can endure the shock of impulse interference, but will collapse when the impulse interference power reaches to a strong level ($\Gamma = 20$ dB in this research).

In the tests with RS(31, 21)+CC(1/2) coded packet, $\Gamma = 0$ dB was not considered for Figure 11 (a), since in this case PER is very low, in the order of $10^{-7}$. Instead, $\Gamma = 6.02$ dB is used in Figure 11 (a), while in Figure 11 (b), (c) and (d), $\Gamma$ with the same values are selected as in Figure 10 (b), (c) and (d). Comparing PER in Figure 10 and 11 the following is concluded: i) At low power ratio $\Gamma$ (< 6.02 dB), the RS(31, 21)+CC(1/2) concatenated code is superior to CC(1/2). This is also consistent with the results that shown in Figure 9 (a) and (c). ii) As the value of $\Gamma$ increases (6.02~20 dB), the RS(31, 21)+CC(1/2) concatenated code loses its superiority over the CC(1/2). Since RS(31, 21)+CC(1/2) is a concatenated coding scheme with a lower code rate, the coded packet is longer. Hence there may be more bits are corrupted by the impulse noise when the packet is transmitted through the wireless channel.

Above analysis implies the correlation between the PER performance and the number of impulse in each data OFDM frame. Taking the S-random interleaver as an example, the comparison of PER versus the number of impulse in one frame is presented in Figure 12. For the CC(1/2) with 74.8 us and the RS(31, 21)+CC(1/2) with 108.8 us data OFDM frame duration, respectively. As can expect is that the PER drops with the increase of impulse number, the higher the impulse noise energy, the faster degradation of the reliability. As the interleaved coded transmission scheme can tolerate weak impulse noise interference, this advantage no longer exists as the impulse noise energy increase, especially for the lower code rated RS(31, 21)+CC(1/2). The rationale behind this phenomenon is that RS(31, 21)+CC(1/2) is a concatenated code. For a 100-bit packet, the packet length is 155 bits after RS(31, 21) encoding. Consequently, the overall packet length is 310 bits after the CC(1=2) encoding, versus to the 200 bits length when only CC(1/2) is employed. From the statistical point of view, we assume the same bit error rate for CC(1/2) coded packet. The 155-bit RS codewords include more error bits than the 100-bit packet after Viterbi decoding, PER of the 155-bit RS codewords is higher than the 100-bit message packet. So, improved PER can be achieved after the RS decoding for weak impulse noise. But in case of strong impulse noise interference, the RS decoder cannot correct the error bits and thus worse PER is expected.

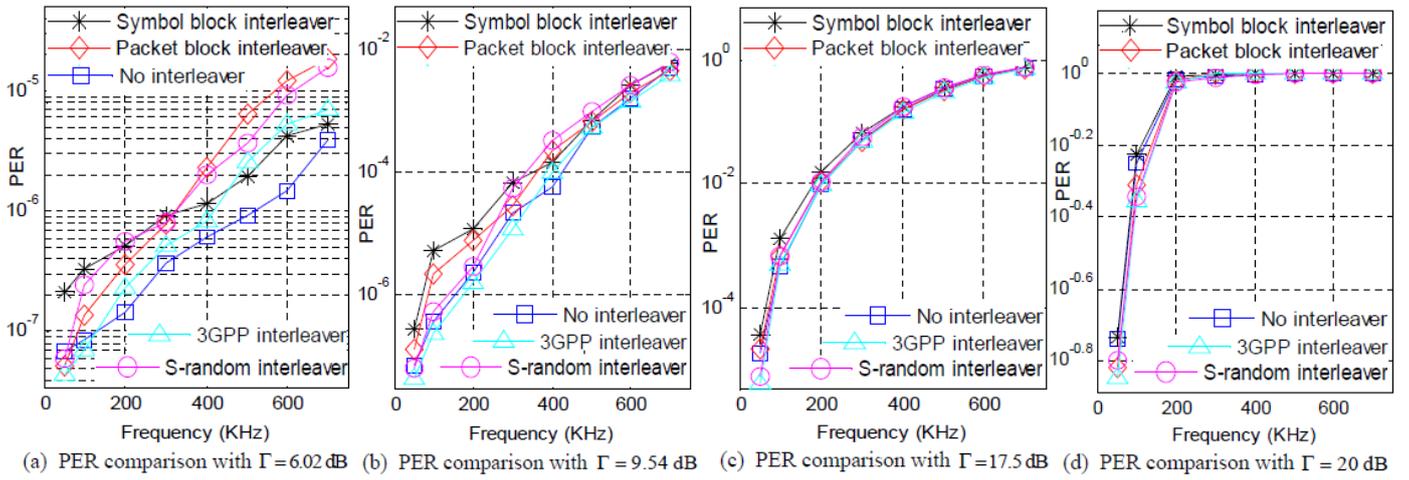

Fig. 11. PER comparison of impulse interference with different power ratio $\Gamma$ for RS(31, 21)+CC(1/2) concatenated code.

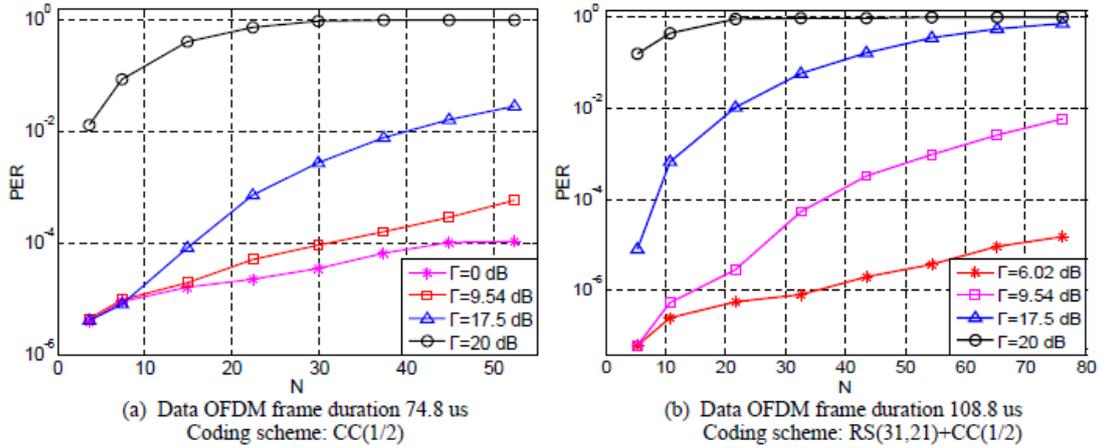

Fig. 12. PER comparison of impulse number (N) in each data OFDM frame using S-random interleaver.

Based on the above experimental results, the PER curves degraded gradually when $\Gamma \leq 17.5$ dB, which shows the impulse noise power is in the linear input range of the receiver USRP radio frequency (RF) front-end. As $\Gamma$ is increased to 20 dB, the impulse noise peak power is 100 times the data OFDM frame transmission power, the PER almost reaches to 1 as the impulse frequency increases. Consequently, we suppose this level of impulse noise power represents the most severe case that the USRP-PC hardware platform can withstand. In real industrial environments, there may exist more powerful impulse interference, for example 25 dB malicious jamming attack, that can saturate the receiver USRP RF front-end. Restricted by the nonlinear effects, samples of the impulse noise in the receiver USRP are similar to the case of 20 dB impulse noise. The WirelessHP system only has one OFDM symbol as the preamble and is less robust to the high-energy impulse surge. Functionalities, such as synchronization, channel estimation, frequency offset and phase correction, will completely collapse to invalid operation, resulting in sudden degradation of the PER performance.

### C. Discussion of interleavers in short-packet transmission

To summarize, when interleaver is adopted for short-packet transmission, impulse interference power, interleaver structure and coding scheme are the three mutually dependent factors that determine PER performance:

1) The impulse interference power and frequency are the crucial factors affecting the reliability. On condition of strong impulse interference and high switching frequency, the improvements of PER performance are small, or degraded when compared with the no interleaver case. Consequently, interleavers are completely unnecessary in extreme scenarios. However, there is still room to use interleavers if the impulse interference power and frequency are within a certain range. Good interleavers, such as the 3GPP and S-random interleavers, can achieve better PER performance by relocating neighbor bits to a longer minimal distance. For coded short-packet transmission, interleavers with longer minimal distance are more suitable for impulse noise deteriorated wireless channels.

2) Interleaver performance tends to depend on the coding schemes. For codes with moderate error correction capability, i.e. the RS(31, k) and RS(31, k)+CC(5/6) codes in Figure 9, the interleavers are more significant to affect the PER performance, as compared to the more powerful CC(1/2) and RS(31, k)+CC(1/2) concatenated codes. Similar phenomenon can also be found in Figure 10 and 11 where CC(1/2) and RS(31, 21)+CC(1/2) concatenated codes are compared, especially in cases of low power ratio $\Gamma$ and high impulse frequency region. As an observed phenomenon, this correlation should be further investigated.

Based on these analyses, impulse interference power and frequency are the main parameters determining the usefulness of interleavers in coded short-packet transmission. Different interleavers deliver different PER performance improvements for different codes. The selection of interleavers also depends on the expected characteristics of interference. In practice, it is of most importance to select a method which is most robust to the widest range of impulse and other interference characteristics.

## VI. CONCLUSIONS

For the WirelessHP system featured with one OFDM symbol as the preamble and 100-bit level short-packet for transmission, this paper investigates the feasibility of applying bit interleaving to its physical layer in a factory environment. Based on the constructed USRP-PC hardware platform for WirelessHP system, we show that bit interleaving in coded short-packet can improve the reliability. Given the same coding scheme, the interleaver structure is essential for improving PER performance. By generating impulse interference with different power levels, we have tested the resilience of interleavers in practical scenarios. We show that the PER improvement depends more on the power and frequency of impulse interference, while the influence of the interleaver structure is less significant. Due to the fact that PER improvement is affected by a series of interdependent factors, a thorough investigation is required to find effective solutions, including the OFDM impulse noise cancellation and smart decoding algorithms, to detect and mitigate the reverse effect of impulse noise for ultra-high reliability industrial wireless control applications.